\documentclass[aps,prd,reprint,onecolumn,notitlepage,superscriptaddress]{revtex4-1}

\usepackage{lipsum}
\usepackage{amsmath}
\usepackage{graphicx}
\usepackage{color}
\usepackage{tikz}
\usepackage{verbatim}

\usepackage{amssymb}
\usepackage{bm}

\newcommand{\rom}[1]{\textup{\uppercase\expandafter{\romannumeral#1}}}

\begin{document}

\title{Modified Gravity Corrections in Fundamental Orbital Frequencies in Kerr Spacetime}
\author{Avani Patel\footnote{email:avani@iiserb.ac.in},
       }
\affiliation{Indian Institute of Science Education and Research Bhopal,\\ Bhopal 462066, India}




\begin{abstract}
As a first step towards the calculation of waveform of Extreme Mass Ratio Inspirals for Modified Gravity theories, we calculate the orbital frequencies of a Small Compact Object inspiralling into a super massive blackhole for a Nonlocal gravity model. The small compact object moves along an orbit which can be approximated to a geodesic of the background spacetime due to large mass ratio of central blackhole to Small Compact Object. In General Relativity, the fundamental orbital frequencies $\Omega_r,\;\Omega_{\theta}$ and $\Omega_{\phi}$ can be calculated by solving geodesic equations of the Kerr metric. If we formulate any modified gravity theory as a small correction in General Relativity then the spacetime metric around a rotating blackhole in that theory can be considered as the Kerr metric with small deformations. This would allow us to calculate fundamental frequencies of geodetic motion of the orbiting object perturbatively i.e. as Kerr frequencies plus small shifts in Kerr frequencies coming from the modified gravity part. Using Action-angle formalism and canonical perturbation theory I calculate the frequency shifts with respect to Kerr frequencies of the orbital motion around a rotating blackhole for RR model of Nonlocal gravity theory. 

\end{abstract}

\maketitle
\section{INTRODUCTION}
Detection and study of Extreme Mass Ratio Inspiral(EMRI) is one of the promising phenomena to probe the nature of gravity and other fundamental physics in era of gravitational wave astronomy\cite{Barack:2018yly,Chua:2018yng,Berry:2019wgg}. EMRIs can be used as standard sirens to measure the Hubble rate of the universe\cite{MacLeod:2007jd}. It is also useful to understand the galactic dynamics and environment of supermassive blackhole(SMBH) at galactic nucleus\cite{Berry:2019wgg}. The first image of supermassive blackhole $M87^*$ taken by Event Horizon Telescope has proved the existence of supermassive blackhole in the centre of a galaxy\cite{Akiyama:2019cqa,Akiyama:2019eap}. 

EMRI is the orbit of stellar mass blackhole or other small compact object(SCO) ($10^0-10^2\;M_{\odot}$)  inspiralling into a supermassive blackhole having mass around $10^5-10^7\;M_{\odot}$. Due to such extreme mass ratio of participating objects, the characteristics of gravitational waves emitted during the process can reflect the spacetime structure of supermassive blackhole with minimum effects of binary dynamics. Before plunging into the central blackhole the small compact object(SCO) completes $10^5-10^6$ orbital cycles having long periodic time. The estimated frequency of the emiited gravitational waves from EMRIs is around $10^{-4}-1$ Hz which makes them primary source of space based interferometer like LISA(Laser Interferometer Space Antenna). Considering different EMRI models, the observational potential of LISA is estimated to be around $100$ per year\cite{Babak:2017tow}. To acquire the information from EMRI gravitational waves it is important to have precise waveform models. 

Because of the small mass ratio (SCO mass $m$ divided by central SMBH mass $M$), perturbative calculation of motion of orbiting object is possible in which leading order is just the geodetic motion of a point particle in central blackhole's spacetime. At higher orders, one has to take into account the effect of SCO's own gravitational field on its motion. This effect is described as self force. In order to obtain the required precision in phase for matched filtering, perturbation theory up to second order i.e. second order self force is necessary in waveform calculation\cite{Hinderer:2008dm}. The first order self force studies have been done extensively in Schwarzschild spacetime but less has been done in Kerr spacetime. In Schwarzschild spacetime, the first order self force has been considered in the calculation of orbital evolution in many works\cite{Warburton:2011fk,Diener:2011cc,Wardell:2014kea,Warburton:2017sxk,Heffernan:2017cad}. Using Near Identity Transformation, the fast method of computing the inspiral trajectory and gravitational waves from EMRIs incorporating self force has been presented in \cite{vandeMeent:2018rms}. The self force has been calculated to first order for an EMRI moving along a generic orbit in Kerr spacetime\cite{vandeMeent:2017bcc}, and there has been extensive work towards a second-order calculation\cite{Galley:2011te,Pound:2014xva,Pound:2014koa,Miller:2016hjv,Pound:2017psq,Moxon:2017ozd}. 
Numerical calculation of the EMRI waveform based on perturbation theory is computationally expensive. Without considering self force some approximate ``Kludge" waveforms are developed to make the modeling more economical. Though such waveforms are not exact but can give the prominent features of the inspiral\cite{Babak:2006uv,Gair:2017ynp}. There are three types of Kludge models prevalent in literature: analytic kludge\cite{Barack:2003fp}, numerical kludge\cite{Babak:2006uv} and augmented analytic kludge\cite{Chua:2017ujo}.

In all above mentioned approaches, the very first step is to solve the geodesic equation. The Kerr geometry enjoys the existence of two Killing vectors and one Killing tensor which gives rise to three constants of motion namely, Energy $E$, axial angular momentum $L_z$ and Carter constant $Q$. The rest mass of the particle $m$ is also a constant of motion. Due to presence of these four constants of motion, geodetic trajectory of a point particle can be fully specified in Kerr spacetime. In case of deviation from the Kerr geometry this aid of symmetry may not be available. In many studies in various directions like tests of no-hair theorem\cite{Vigeland:2009pr}, strong field tests of General Relativity(GR)\cite{Carson:2020dez,Carson:2020iik} and modified gravity theories\cite{Sopuerta:2009iy,Vigeland:2011ji,Mizuno:2018lxz}, the central SMBH spacetime metric is indeed different from the Kerr metric. If the deviation from the Kerr metric is very small compared to the Kerr terms then one can use the canonical perturbation theory to solve the geodesic equations\cite{Vigeland:2009pr}.

The late-time acceleration is one of the outstanding problem in modern cosmology\cite{Nobel2011} which GR has failed to explain. Due to this shortcoming of GR many modified gravity theories are constructed mostly as a small modification to GR. One such modified gravity is Nonlocal gravity which has emerged as an effective candidate for cosmological constant. In a line of research, nonlocal gravity models are constructed with the motivation to write massive term without breaking the gauge invariance of massless theory\cite{Dvali:2006su,Kehagias:2014sda,Maggiore:2014sia}. The mass parameter enters in a theory as a coefficient of nonlocal term. In this work I consider a specific model, called $RR$ model, that has correction term proportional to $R\frac{1}{{\Box}^{2}}R $, in Einstein-Hilbert action, proposed in\cite{Maggiore:2014sia}. In the above study, the static and spherically symmetric metric for $RR$ model is calculated. Considering this static and spherically symmetric solution of $RR$ model, the stationary and axisymmetric solution is calculated in \cite{Kumar:2019uwi}. In the present work, I write the above stationary and axisymmetric metric in the form of Kerr metric plus small deformation which corresponds to nonlocal correction and solving the geodesic equations in the deformed background I calculate the fundamental frequencies of the orbital motion.

The plan of this paper is as follows. In Sec.~\ref{sec:Kerrfreq}, I review the geodesic structure of the Kerr metric and calculation of fundamental orbital frequencies $\Omega_{r,\theta,\phi}$ in the Kerr spacetime. In Sec.~\ref{sec:NLGfreqshifts}, shifts in above fundamental frequencies due to nonlocal correction of RR model are calculated. I introduce RR model of the nonlocal gravity and the rotating blackhole metric for the model in Sec.~\ref{ssec:rotbhNLG} and then in Sec.~\ref{ssec:comp}, the computations of the shifts $\delta \Omega_{r,\theta,\phi}$ in Kerr frequencies are done using canonical perturbation theory. I finally discuss our results and conclude the paper in Sec.~\ref{sec:discussions}.


\section{Fundamental Orbital Frequencies in Kerr Spacetime}\label{sec:Kerrfreq}

The equation of motion of a point mass $m$ moving along the geodesic of any metric $g_{\alpha\beta}$ is given by geodesic equation $\dot{u^{\alpha}} + \Gamma_{\rho\sigma}^{\alpha}u^{\rho}u^{\sigma}=0$, where $u^{\alpha}$ is the four-velocity of the point mass, overdot denotes the derivative w.r.t. proper time $\tau$ and $\Gamma_{\rho\sigma}^{\alpha}$ is the Levi-Civita connection associated with the metric $g_{\alpha\beta}$. These four coupled equations can be very complicated to solve for an arbitrary metric. The natural way to proceed is to look for the symmetry of the spacetime. Let us first write the Kerr metric in Boyer-Lindquist coordinates $(t, r, \theta, \phi)$
\begin{eqnarray}
 ds^2 = -\left[1-\frac{2Mr}{\Sigma}\right]\;dt^2- \left[\frac{4Mr}{\Sigma}a\sin^2\theta \right]\;dt\;d\phi + \frac{\Sigma}{\Delta} dr^2+ \Sigma\; d\theta^2+ \left[\sin^2\theta \left\{\Sigma +\right.\right.\nonumber\\ \left.\left.\left(1+\frac{2Mr}{\Sigma}\right)a^2\sin^2\theta\right\}\right]\;d\phi^2, 
\label{kerrmetric}         
\end{eqnarray}
where $\Delta \equiv r^2 +a^2 - 2Mr$ and $\Sigma  \equiv  r^{2} + a^{2}\cos ^{2}\theta$ and $a$ and $M$ are spin parameter and mass of the central blackhole respectively.    
It is apparent that the Kerr metric is independent of coordinates $t$ and $\phi$ and therefore the Kerr spacetime possess two Killing vectors, a timelike one $\xi^{\alpha}_{(t)}=\delta^{\alpha}_{t}$ and a spacelike one $\xi^{\alpha}_{(\phi)}=\delta^{\alpha}_{\phi}$. This symmetry gives two obvious constants of motion namely orbital energy $E=-p_t$ and z-component of orbital angular momentum $L_z=p_{\phi}$. Here, $p_{\alpha}\equiv m\; g_{\alpha\beta}\;u^{\beta}$ is the four momentum of the test particle and the Hamiltonian of the system can be written as $H = \frac{1}{2} g^{\alpha\beta}\;p_{\alpha}\;p_{\beta}$. 

The remaining symmetry is not very obvious. With the help of Hamilton-Jacobi method, B.Carter has calculated the third constant of motion known as Carter constant\cite{Carter:1968rr}. Suppose $S$ is a generating function of the canonical transformation from coordinates and momenta $(q_i,p_i)$ to a new set of constant quantities determined by eight initial values, where $q_i=(t, r, \theta, \phi)$ in our case. Using the Hamilton-Jacobi equation and writing $S$ in the separable form where each term depends only on one coordinate $q_i$, one can calculate the Carter constant denoted as $Q\equiv p_{\theta}^2 + \cos^2\theta[a^2(m^2-E^2)+\csc^2\theta L_z^2]$\cite{Carter:1968rr}. This corresponds to the existence of a Killing tensor $K_{\alpha\beta} = 2\Sigma l_{(\alpha}n_{\beta)} + r^2g_{\alpha\beta}$. The fourth trivial constant is the square of rest mass of the test particle $m^2$. Separation of variables can enable us to write the Hamilton-Jacobi(H-J) equation as four restricted H-J equations for each of the four coordinates $t, r, \theta, \phi$\cite{classicalmec}. The integrated form of the geodesic equations can be obtained by taking partial derivatives of $S$ w.r.t. constants $E,L_z,Q$ and $m^2$ and equating them to zero\cite{Carter:1968rr}. Listing out four constants of motion, we have
    \begin{equation}
   E = -m u_t =m (-g_{tt}\dot t - g_{t\phi}\dot{\phi}) \label{19}
    \end{equation}
    \begin{equation}
  L_z = m u_{\phi} = m(g_{\phi t}\dot t + g_{\phi\phi}\dot{\phi}) \label{20} 
    \end{equation}
    \begin{equation}
  Q = m^2(K_{tt} \dot t^2 + 2K_{tr}\dot t \dot r) - (L_z - aE)^2 \label{21}
    \end{equation}
    \begin{equation}
  -m^2 = m^2\;g_{\alpha\beta} u^{\alpha}u^{\beta} = m^2\;(g_{tt}\dot t^2 + 2g_{t\phi} \dot t \dot{\phi} + g_{rr}\dot r^2 + g_{\theta\theta}\dot{\theta}^2 + g_{\phi\phi}\dot{\phi}^2 ).\label{22}
    \end{equation}
Rearranging above equations we end up with four geodesic equations for the Kerr metric\cite{Vigeland:2009pr}
\begin{eqnarray}
m^2\Sigma^2\left(\frac{dr}{d\tau}\right)^2= [(r^2+a^2)E-a L_z]^2-\Delta[m^2r^2+(L_z-aE)^2+Q]\equiv R(r)\nonumber\\
m^2\Sigma^2\left(\frac{d\theta}{d\tau}\right)^2= Q-cot^2\theta L_z^2 - a^2 cos^2\theta(m^2-E^2)\equiv \Theta(\theta)\nonumber\\
m\Sigma\left(\frac{d\phi}{d\tau}\right)= csc^2\theta L_z + aE\left(\frac{r^2+a^2}{\Delta}-1\right)-\frac{a^2L_z}{\Delta}\equiv\Phi(r, \theta)\nonumber\\
m\Sigma\left(\frac{dt}{d\tau}\right)=E\left[\frac{(r^2+a^2)^2}{\Delta}-a^2sin^2\theta\right]+aL_z\left(1-\frac{r^2+a^2}{\Delta}\right) \equiv T(r, \theta),
\label{Kerr_geodesic}    
\end{eqnarray}
The radial coordinate $r$ and the polar coordinate $\theta$ are not very appropriate for orbital motion because $\frac{dr}{d\tau}$ and $\frac{d\theta}{d\tau}$ become zero and change their sign during the orbital motion of the point particle. Inspired from the studies of classical planetary motion, we reparametrize $r$ and $\theta$ as
\begin{equation}
    r\equiv \frac{pM}{1+e\;\cos{\psi_r}};\;\;\;\cos{\theta}\equiv\cos{\theta_{min}}\;\cos{\psi_{\theta}},
    \label{reparam}
\end{equation}
where $p$ is semilatus rectum and $e$ is the eccentricity of the orbit. $\theta_{min}$ and $\pi-\theta_{min}$ are the turning points of the polar motion where $\Theta(\pm\cos\theta_{min})=0$ and angle $\psi_{\theta}$ takes values $0$ and $\pi$ respectively. The turning points of the radial motion are given by periapsis $r=r_p=\frac{pM}{1+e}$ and apoapsis $r=r_a=\frac{pM}{1-e}$ where $R(r_p)=R(r_a)=0$ and angle $\psi_r$ takes values $0$ and $\pi$ respectively. 

In case of periodic motion, it is useful to transform to action-angle variables instead of arbitrary constant quantities. For the system where Hamiltonian is conserved and separation of variables is possible, the generating function $S$ is Hamilton's characteristic function $W$ and it can be written as $W=\sum_i W_i$. Since $W$ is the generating function of canonical transformation, old momenta of the system can be expressed as $p_i=\frac{\partial W}{\partial q_i}$. The action variables are defined as $\hat{J}_i\equiv \oint p_i dq_i$ and the corresponding position coordinates are called angles $\hat{w}_i$ which can be calculated from the expression $\hat{w}_i=\frac{\partial W}{\partial \hat{J}_i}$. For later convenience, we denote action and angle variables for the Kerr spacetime with overhat like $\hat{J}_i$ and $\hat{w}_i$. The definition of the action variables $\hat{J}_i$'s enable one to write the Hamiltonian in terms of $\hat{J}_i$'s. To calculate the frequencies of the orbits one need to write the Hamiltonian $H$ in terms of $\hat{J}_i$'s and calculate  $m\hat{\omega}^{i}=\frac{\partial H^{aa}(J)}{\partial \hat{J}_i}$, where superscript `aa' denotes the Hamiltonian written in terms of action-angle variables\cite{classicalmec, Vigeland:2009pr}. Following the notation of \cite{Vigeland:2009pr}, let us write the angle variables for our case
\begin{eqnarray}
\hat{J}_r \equiv \frac{1}{2\pi} \oint p_r dr=\frac{1}{\pi}\int_{r_p}^{r_a} \frac{\sqrt{R}}{\Delta}dr;\nonumber \\
\hat{J}_{\theta} \equiv \frac{1}{2\pi} \oint p_{\theta} d\theta=\frac{2}{\pi}\int_{\theta_{min}}^{\pi/2} \sqrt{\Theta(\theta)}d\theta;\nonumber \\
\hat{J}_{\phi} \equiv \frac{1}{2\pi} \oint p_{\phi} d\phi=L_z;\nonumber\\
\hat{J}_t = -E.\label{actionvariables}
\end{eqnarray}
It can be seen that the first two integrals in above Eq.\eqref{actionvariables} are not analytically solvable and therefore writing Hamiltonian $H$ in terms of $J_i$'s is not trivial. However, the derivatives $\frac{\partial H^{aa}(J)}{\partial \hat{J}_i}$ can be calculated without knowing the functional form of $H^{aa}(J)$ (please refer the appendix A of \cite{Schmidt:2002qk}). The orbital frequencies $\hat{\omega}^r, \hat{\omega}^{\theta}$ and $\hat{\omega}^{\phi}$ for the Kerr metric are computed in \cite{Vigeland:2009pr, Schmidt:2002qk}. These are called fundamental frequencies of the Kerr spacetime because they characterize the fundamental properties of the orbital motion in the Kerr spacetime. The fundamental frequencies $\hat{\omega}^r, \hat{\omega}^{\theta}$ and $\hat{\omega}^{\phi}$ are associated with proper time which does not have much value from measurement point of view. The observational frequencies which associate with coordinate time are given by $\hat{\Omega}^i=\frac{\hat{\omega}^i}{\hat{\omega}^t}$.


\section{Shifts in Fundamental Orbital Frequencies due to Nonlocal Gravity}\label{sec:NLGfreqshifts}
Most modified gravity models are studied as small correction to GR so that they uphold all experimentally proved results of GR. Though the correction term is weak in comparison with GR it is strong enough to show its effect at the appropriate scale, generally cosmological scale. In this section, our intention is to calculate the effect of such small correction term in RR model of nonlocal gravity in the surrounding region of an SMBH. In particular, we want to calculate the shifts in orbital fundamental frequencies in a rotating blackhole spacetime caused by nonlocal correction to GR.

\subsection{Rotating Blackhole Spacetime for RR Model}\label{ssec:rotbhNLG}
Let us first write the action for the RR model of nonlocal gravity,
\begin{equation}
S = \frac{1}{2\kappa^{2}}\int d^{4}x \sqrt{-g}\Big[ R + \frac{\mu^{2}}{3} R\frac{1}{\Box^2}R\Big] + \mathcal{L}_{m},  \label{action}
\end{equation}
where $\mu^{-1}$ is the length scale associated with nonlocal correction to the Einstein-Hilbert(EH) action and $\mathcal{L}_{m}$ is the Lagrangian for the usual matter. Introducing two scalars
\begin{equation}
 U=-\Box^{-1}R,\;\;\; S=-\Box^{-1}U;\label{auxfields}   
\end{equation}
 and varying the action \eqref{action} w.r.t. $g_{\mu\nu}$ we obtain the equation of motion
\begin{equation}
\kappa^{2} T_{\alpha \beta}  =  G_{\alpha \beta} - \frac{\mu^{2}}{3}\Big\{ 2\Big(G_{\alpha
 \beta} - \nabla_{\alpha}\nabla_{\beta} + g_{\alpha \beta} \Box \Big)S  + g_{\alpha \beta} \nabla^{\gamma}U \nabla_{\gamma}S   - \nabla_{(\alpha}U \nabla_{\beta)}S  -\frac{1}{2}g_{\alpha \beta} U^{2} \Big\},
 \label{fieldeq}
 \end{equation}
along with Eq.\eqref{auxfields}. $G_{\alpha \beta}$ is the Einstein tensor and $T_{\alpha \beta}$ is the energy-momentum tensor of the matter. The fields $U$ and $S$ are auxiliary fields and do not represent radiative degrees of freedom. For more details on this model and other nonlocal gravity models the reader can refer \cite{Maggiore:2014sia,Woodard:2014iga,Calcagni:2010ab,Dirian:2014xoa,Kehagias:2014sda,Nersisyan:2016hjh,Fernandes:2017vvo,Kumar:2018pkb,Kumar:2018chy,Tian:2018bmn,Tian:2019bla,Chen:2019wlu,Joshi:2019cyk}. 

The spherically symmetric and static metric for the above action has been calculated in \cite{Maggiore:2014sia} in the region $r_s\lesssim r\ll m^{-1}$, where $r_s\equiv 2M$ is the Schwartzschild radius and $r$ is the radial coordinate of the coordinate system $(t, r, \theta, \phi)$ used. Since finding the exact solution of the field equations\eqref{auxfields} and \eqref{fieldeq} is very difficult, even in case of spherical symmetry, the above metric is given as first order correction to usual Schwartzschild background of GR. In \cite{Kumar:2019uwi}, the stationary and axisymmetric solution, which describes the spacetime around a slowly rotating blackhole, is computed considering the perturbation up to second order in rotation parameter $a$ and coupling parameter $\mu^2$ on top of the spherically symmetric and static solution shown in \cite{Maggiore:2014sia}. Let us write the metric computed in \cite{Kumar:2019uwi} for the spacetime around rotating blackhole in RR model as in the form of $g_{\alpha\beta}=g_{\alpha\beta}^{Kerr}+b_{\alpha\beta}$,
\begin{eqnarray}
ds^2 = -\left[1-\frac{2Mr}{\Sigma}-\frac{2Mr\mu^2}{6}\right]\;dt^2- \left[\frac{4Mr}{\Sigma}a\sin^2\theta +C^2 a\sin^2\theta\left(- \frac{1066}{534}\left(\frac{2M}{r}\right)\right.\right.\nonumber\\
\left.\left. - \frac{1}{6}\left(\frac{2M}{r}\right)^2- \frac{2}{9}\left(\frac{2M}{r}\right) \ln\left(\frac{2M}{r}\right)\right)\right]\;dt\;d\phi + \left[\frac{\Sigma}{\Delta} + \frac{2Mr\mu^2}{6\left(1-\frac{2M}{r}\right)^2}\right]dr^2 \nonumber \\
 + \Sigma \;d\theta^2+ \left[\sin^2\theta \left\{\Sigma + \left(1+\frac{2Mr}{\Sigma}\right)a^2\sin^2\theta\right\}\right]\;d\phi^2,    
\label{nlgmetric}    
\end{eqnarray}
where $C\equiv 2M\mu$. As done in \cite{Vigeland:2011ji}, we can map the deformations in Kerr blackhole sourced by modified gravity with bumpy blackhole structure. Unlike \cite{Vigeland:2009pr}, here, deformations in Kerr geometry are non-GR in nature but geodesic motion is still valid upto leading order in mass-ratio\cite{Vigeland:2011ji,Gralla:2010cd}. Using the same arguments as shown in Sec.IIIB in \cite{Sopuerta:2009iy}, one can also show that the trajectory of the point particle is described by the geodesic equation as GR also in our case. Therefore, it is harmless to consider the $b_{\alpha\beta}$ terms in Eq.\eqref{nlgmetric} as bumps on Kerr blackhole and study the geodetic motion following the same approach taken for bumpy Kerr blackhole in \cite{Vigeland:2009pr}. For the metric \eqref{nlgmetric}, the deformations in Kerr metric are parametrized by one parameter which is $\mu$ and as $\mu\rightarrow 0$ our metric smoothly reduces to the Kerr metric. In the next subsection, we will treat $b_{\alpha\beta}$ as the perturbation in the Kerr metric and calculate the shifts in Kerr frequencies using canonical perturbation theory. 


\subsection{Computation of Shifts Using Canonical Perturbation Theory}\label{ssec:comp}

Since the nonlocal correction term added to Einstein-Hilbert(EH) action is very small compared to EH term we can use canonical perturbation theory to solve the geodesic equations of nearly Kerr-like blackhole of nonlocal gravity model. The Hamiltonian for the system of SCO moving around an SMBH is the same as written in Sec.\ref{sec:Kerrfreq}, $
H = \frac{1}{2} g^{\alpha\beta}\;p_{\alpha}\;p_{\beta}$. But now the metric is given by $g_{\alpha\beta}=g_{\alpha\beta}^{Kerr}+b_{\alpha\beta}$. If we consider $b_{\alpha\beta}$ as the perturbation in the background Kerr spacetime then lowering and rising of any tensor can be done with the background metric $g_{\alpha\beta}^{Kerr}$. Thus the contravariant metric is given by $g^{\alpha\beta}=g_{Kerr}^{\alpha\beta}-g_{Kerr}^{\alpha\mu}\;g_{Kerr}^{\beta\nu}\;b_{\mu\nu}$. Therefore, we can write the Hamiltonian as
\begin{eqnarray}
H = H_{Kerr} + H_{NL};\label{pertH}
\end{eqnarray}
where, $H_{NL} = -(1/2)b^{\alpha\beta}p_{\alpha}p_{\beta}$ and `NL' stands for the nonlocal correction. The $H_{NL}$ is the perturbation to the original Hamiltonian $H_{Kerr}$. 

The canonical properties of the canonical transformation $(q_i,p_i)\rightarrow (\hat w_i,\hat J_i)$ by the generating function $W(q_i, J_i)$ remains valid also in case of perturbed $H$ and so the $(\hat w_i, \hat J_i)$ remain valid canonical set of variables. If we express the total Hamiltonian $H$ in terms of $\hat w_i$ and $\hat J_i$ then it can be expanded in power of small perturbation parameter. Writing the Hamilton-Jacobi equation for the generating function which generates the transformation from $(\hat{w}_i,\hat{J}_i)$ to $({w}_i,J_i)$ such that all $J_i$'s are constants and $w_i$'s are linear functions of time, expanding it on both sides in power of perturbation parameter and equating the terms having same power of perturbation parameter one leads to the computation of perturbation in the frequency $\hat{\omega}^i$ resulting from the presence of the perturbation $H_{NL}$ in the total Hamiltonian $H$. The reader can refer \cite{classicalmec} for detailed formalism of canonical perturbation theory. Here, we directly write the formula for the orbital frequencies associated with the perturbed system as
\begin{eqnarray}
\omega^i=\hat{\omega}^i+\delta\omega^i;\nonumber\\
m\;\delta\omega^i=\frac{\partial\langle H_{NL}\rangle}{\partial\hat{J}_i},\label{freqshift}
\end{eqnarray}
where, $\langle H_{NL}\rangle$ shows the averaged Hamiltonian $H_{NL}$ over a period of the orbit in background spacetime. Computation of frequencies comprises of three steps : (i) Computation of $H_{NL}$ (ii) Averaging of $H_{NL}$ and (iii) differentiation of $\langle H_{NL}\rangle $ w.r.t. $\hat J_i$. We will proceed understanding each of these three steps. From Eq.\eqref{pertH}, $H_{NL}$ can be written as
\begin{equation}
H_{NL}=-\frac{1}{2}b_{\mu\nu}\left(g^{\alpha\mu}_{Kerr}\;p_{\alpha}\right)\;\left(g^{\beta\nu}_{Kerr}\;p_{\beta}\right)
\label{pertH1}
\end{equation}
Since lowering and rising of any tensor is done by Kerr metric, using Eq.\eqref{Kerr_geodesic} and the nonzero components $b_{tt}, b_{rr}$ and $b_{t\phi}$ of the metric \eqref{nlgmetric}, we obtain
\small
\begin{equation}
H_{NL}(r, \theta)=-\frac{1}{2\Sigma^2}\left[b_{tt}\left(T(r,\theta)\right)^2+b_{rr}\left(R(r)\right)+2b_{t\phi}\left(T(r, \theta)\right)\left(\Phi(r, \theta)\right)\right]
\label{pertH2}
\end{equation}
\normalsize 
One can note that $H_{NL}$ is the non-saparable function of $r$ and $\theta$ due to presence of $\Sigma$ in denominator in Eq.\eqref{pertH2}. The time average of $H_{NL}$ over an orbit can be done as
\begin{equation}
\langle H_{NL}(r(t),\theta(t))\rangle=\lim_{t_p\rightarrow\infty}\frac{1}{2t_p}\int_{-t_p}^{t_p}H_{NL}\; dt.
\label{pertH3}
\end{equation}
Here, $t_p$ is the periodic time of the orbit. Since the radial and polar motion are nonseparable in terms of coordinate time $t$ the computation of above integration is difficult. If we define another time variable $\lambda$ such that $d\lambda=d\tau/\Sigma$ the equations of motion in Eq.\eqref{Kerr_geodesic} becomes
\begin{eqnarray}
m^2\left(\frac{dr}{d\lambda}\right)^2 =  R(r)\nonumber\\
m^2\left(\frac{d\theta}{d\lambda}\right)^2 =  \Theta(\theta)\nonumber\\
m\left(\frac{d\phi}{d\lambda}\right) = \Phi(r, \theta)\nonumber\\
m\left(\frac{dt}{d\lambda}\right) = T(r, \theta).
\label{Kerr_geodesic_lambda} 
\end{eqnarray}
It is apparent that the radial motion and polar motion is now separable and the calculation of fundamental frequencies $\Upsilon^{r,\theta}$ for $r$ and $\theta$ -motion with respect to this new time $\lambda$ is straightforward. The fundamental periods corresponding to radial and polar motion are given by
\begin{eqnarray}
\Lambda^{r}= \int_0^{\Lambda^r} d\lambda = 2 \int_{r_p}^{r_a}\frac{dr}{\sqrt{R(r)}};\nonumber \\
\Lambda^{\theta}= \int_0^{\Lambda^{\theta}} d\lambda = 4 \int_{\theta_{min}}^{\pi/2}\frac{d\theta}{\sqrt{\Theta(\theta)}}.
\label{Kerr_period_lambda}
\end{eqnarray}
and then fundamental frequencies are $\Upsilon^{r,\theta}=2\pi/\Lambda^{r,\theta}$ and the corresponding angles are $w^{r,\theta}=\Upsilon^{r,\theta}\lambda$. The time variable $\lambda$ for geodetic motion in Kerr spacetime was first introduced by Mino and therefore known as ``Mino" time in literature\cite{Mino:2003yg}. It is important to notice that the relation between two time variables, the coordinate time $t$ and the ``Mino" time $\lambda$, is not linear since $T(r,\theta)$ is the function of spacetime coordinates $r$ and $\theta$. Therefore, in order to relate two time periods $\tau^i(=2\pi/\hat\omega^i)$ and $\Lambda^i$ one can use averaged $T(r,\theta)$ over some interval $\Delta\lambda$ which is denoted as $\langle T\rangle_{\lambda}$. Thus $\tau^i=\langle T\rangle_{\lambda}\;\Lambda^i$. According to \cite{Fujita:2009bp}, this is nothing but the frequency of t-motion associated with ``Mino" time. We can show
\begin{eqnarray}
    \Upsilon^t &=& \langle T(r,\theta)\rangle_{\lambda}\nonumber\\
               &=& \langle T_r(r)\rangle_{\lambda} + \langle T_{\theta}(\theta)\rangle_{\lambda} + aL_z\nonumber\\
               &=& \frac{2}{\Lambda^r} \int_{\lambda(r_p)}^{\lambda(r_a)}  T_r(r)\;d\lambda+ \frac{4}{\Lambda^{\theta}} \int_{\lambda(\theta_{min})}^{\lambda(\pi/2)}  T_{\theta}(\theta)\;d\lambda + aL_z\nonumber\\
               &=& \frac{1}{2\pi} \int_0^{2\pi} T_r(r)\; dw^r + \frac{1}{2\pi} \int_0^{2\pi} T_{\theta}(\theta)\; dw^{\theta} + aL_z\nonumber\\
               &=&\frac{1}{(2\pi)^2}\int_0^{2\pi}dw^r\int_0^{2\pi}dw^{\theta}\;T(r(w^r),\theta(w^{\theta}))
  \label{Kerr_tfreq_lambda}
\end{eqnarray}
In other words, $\Upsilon^t$ is average rate at which $t$ accumulates as a function of $\lambda$\cite{Drasco:2003ky}. In fact, average rate at which any arbitrary function $f(r,\theta)$ accumulates as a function of $\lambda$ can be given by
\begin{equation}
 \langle f(r,\theta)\rangle_{\lambda}=\frac{1}{(2\pi)^2}\int_0^{2\pi}dw^r\int_0^{2\pi}dw^{\theta}\;f(r(w^r),\theta(w^{\theta})).  
 \label{arbf}
\end{equation}
Now we can rewrite Eq.\eqref{pertH3} as
\begin{equation}
\langle H_{NL}(r(t),\theta(t))\rangle=\frac{\int_{\lambda(-t_p)}^{\lambda(t_p)}H_{NL}\;\frac{dt}{d\lambda}d\lambda}{\int_{\lambda(-t_p)}^{\lambda(t_p)}\frac{dt}{d\lambda}d\lambda}.
\label{pertH31}
\end{equation}
The r.h.s. of the above equation can be written as the ratio of two averaged quantities as $\langle H_{NL}T \rangle_{\lambda}/\langle T \rangle_{\lambda}$.
\begin{equation}
\langle H_{NL}(r,\theta)\rangle=\frac{\frac{1}{(2\pi)^2}\int_0^{2\pi}dw^r\int_0^{2\pi}dw^{\theta}\;H_{NL}(r(w^r),\theta(w^{\theta}))\;T(r(w^r),\theta(w^{\theta}))}{\frac{1}{(2\pi)^2}\int_0^{2\pi}dw^r\int_0^{2\pi}dw^{\theta}\;T(r(w^r),\theta(w^{\theta}))}
\label{pertH4}
\end{equation}

For the practical purpose of computation of $\langle H_{NL}\rangle$, we write every thing in r.h.s. of Eq.\eqref{pertH4} in terms of $p, e, \theta_{min}, \psi_r$ and $\psi_{\theta}$ as defined in \eqref{reparam}. The integrand in the numerator of the Eq.\eqref{pertH4}
\begin{equation}
H_{NL}T=-\frac{1}{2\Sigma^2}\left[b_{tt}\left(T(r,\theta)\right)^3+b_{rr}\left(R(r)\right)\left(T(r,\theta)\right)+2b_{t\phi}\left(T(r, \theta)\right)^2\left(\Phi(r, \theta)\right)\right]
\label{H1T}
\end{equation}
can be completely written in terms of $p, e, \theta_{min}, \psi_r$ and $\psi_{\theta}$ as the constants of motion $E, L_z$ and $Q$ are also expressed in terms of them. At the turning points of $r$ and $\theta$ -motion, the equations $R(r_p)=R(r_a)=0$ and $\Theta(\theta_{min})=0$ are satisfied. The system of these three equations enables one to write three constants of motion in terms of $p, e$ and $\theta_{min}$ as shown in appendix of \cite{Schmidt:2002qk} in details. Differentials of angles $w^r$ and $w^{\theta}$ are given by
\begin{eqnarray}
dw^r = \frac{2\pi}{\Lambda^r}\frac{1}{\sqrt{R}}\frac{pM}{(1+e\cos{\psi_r})^2}e\sin{\psi_r}\;d\psi_r;\nonumber\\
dw^{\theta}= \frac{2\pi}{\Lambda^{\theta}}\frac{1}{\sqrt{\Theta}}\frac{\cos{\theta_{min}}\sin{\psi_{\theta}}}{\sqrt{1-(\cos{\theta_{min}}\cos{\psi_{\theta}})^2}}d\psi_{\theta}.
\label{dwrdwtheta}
\end{eqnarray}
The integration limits of variables $\psi_r$ and $\psi_{\theta}$ are $0$ to $2\pi$. After completing the above calculation we end up with $\langle H_{NL} \rangle$ expressed in terms of $m, p, e$ and $\theta_{min}$. Likewise we have actions $\hat J_t, \hat J_r, \hat J_{\theta}$ and $\hat J_{\phi}$ written in terms of $m, p, e$ and $\theta_{min}$ in Eq.\eqref{actionvariables}. We take the similar approach as directed in \cite{Vigeland:2009pr} to calculate $\partial \langle H_{NL} \rangle/\partial \hat J_i$. We first define two matrices \textbf{J} and \textbf{K} as\\
\[
   \textbf{J}\equiv
  \left[ {\begin{array}{cccc}
   \frac{\partial \langle H_{NL}\rangle}{\partial m} & \frac{\partial \langle H_{NL}\rangle}{\partial p} & \frac{\partial \langle H_{NL}\rangle}{\partial e} & \frac{\partial \langle H_{NL}\rangle}{\partial\theta_{min}} \\
  \end{array} } \right]
\]
\[
   \textbf{K}\equiv
  \left[ {\begin{array}{cccc}
   \frac{\partial \hat J_t}{\partial m} & \frac{\partial \hat J_t}{\partial p} & \frac{\partial \hat J_t}{\partial e} & \frac{\partial \hat J_t}{\partial\theta_{min}} \\
   \frac{\partial \hat J_r}{\partial m} & \frac{\partial \hat J_r}{\partial p} & \frac{\partial \hat J_r}{\partial e} & \frac{\partial \hat J_r}{\partial\theta_{min}} \\
   \frac{\partial \hat J_{\theta}}{\partial m} & \frac{\partial \hat J_{\theta}}{\partial p} & \frac{\partial \hat J_{\theta}}{\partial e} & \frac{\partial \hat J_{\theta}}{\partial\theta_{min}} \\
   \frac{\partial \hat J_{\phi}}{\partial m} & \frac{\partial \hat J_{\phi}}{\partial p} & \frac{\partial \hat J_{\phi}}{\partial e} & \frac{\partial \hat J_{\phi}}{\partial\theta_{min}} \\
  \end{array} } \right]
\]
From Eq.\eqref{freqshift}, the shift in fundamental frequency, using the chain rule, can be calculated as
\begin{equation}
    \delta \omega^i = \frac{1}{m}\left(\textbf{J}\textbf{K}^{-1}\right)_i.
    \label{freqshift1}
\end{equation}
The shift in observable frequencies are given by
\begin{equation}
\delta\Omega^i = \frac{\delta \omega^i}{\hat\omega^t}-\frac{\hat\omega^i\delta\omega^t}{(\hat\omega^t)^2}
\label{obsfreqshift}
\end{equation}
 

\section{Discussion and Results}\label{sec:discussions}
\begin{figure}[!h]
  \centering
   $
   \begin{array}{c c}
   \includegraphics[width=0.45\textwidth]{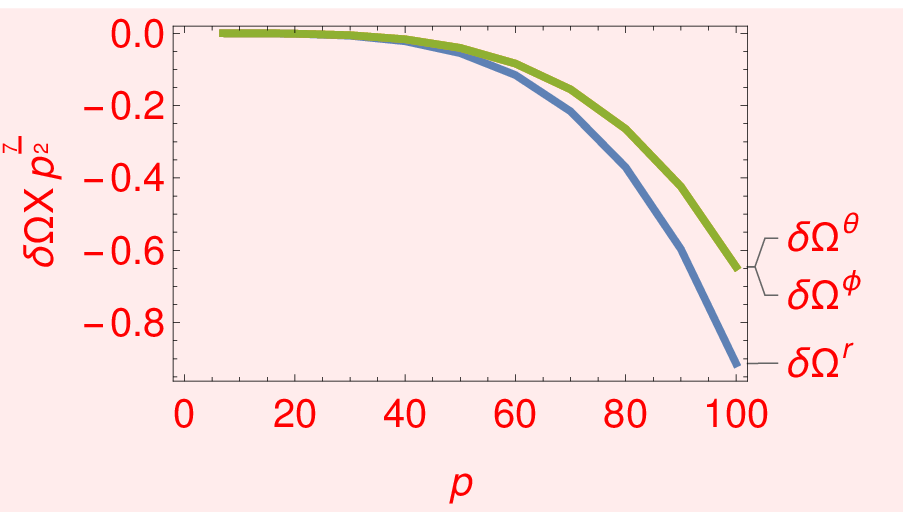} &
   \includegraphics[width=0.45\textwidth]{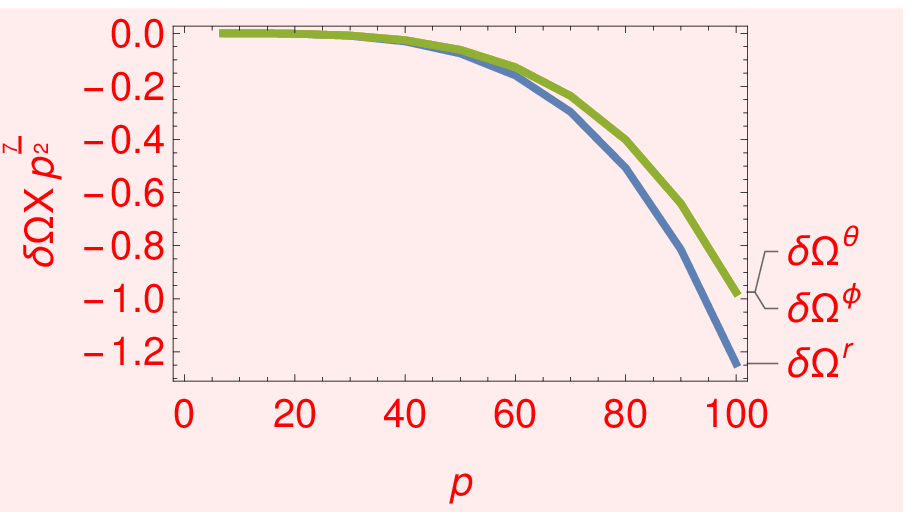} 
   \end{array}
   $ 
  \caption{shift in observable orbital frequencies due to nonlocal correction for $\mu=0.0001$, $\theta_{min}=\pi/3$, $e=0.3$(left) and $e=0.7$(right).}
  \label{fig:orbitalfreq1}
\end{figure}

I numerically calculate the shift in fundamental frequencies from the Kerr frequencies due to nonlocal correction of $\mu^2R\frac{1}{\Box^2}R$ which is shown in plot of Fig.[\ref{fig:orbitalfreq1}]. 
We can see that the shift due to nonlocal gravity in fundamental frequencies of Kerr spacetime gets larger as we go away from the central blackhole. This implies that the nonlocal correction terms in $tt$ and $rr$-component of the metric\eqref{nlgmetric} contributes more compared to the $t\phi$ component. The observational potential of this shift due to nonlocal gravity is higher in the initial inspiral than in the region deep into the SMBH's field.

\begin{table}
\begin{center}
\begin{tabular}{ |c|c|c| } 
 \hline
 p & $\mu=0.0001$ & $\mu=0.00001$ \\ 
 \hline
 10 & -1.8420 x $10^{-8}$ & -1.8420 x $10^{-10}$ \\ 
 20 & -3.4810 x $10^{-8}$ & -3.4810 x $10^{-10}$ \\ 
 30 & -4.5800 x $10^{-8}$ & -4.5800 x $10^{-10}$ \\ 
 40 & -5.4648 x $10^{-8}$ & -5.4648 x $10^{-10}$ \\ 
 50 & -6.2262 x $10^{-8}$ & -6.2262 x $10^{-10}$ \\ 
 60 & -6.9044 x $10^{-8}$ & -6.9044 x $10^{-10}$ \\ 
 70 & -7.5219 x $10^{-8}$ & -7.5219 x $10^{-10}$ \\ 
 80 & -8.0924 x $10^{-8}$ & -8.0924 x $10^{-10}$ \\ 
 90 & -8.6253 x $10^{-8}$ & -8.6253 x $10^{-10}$ \\ 
 100 & -9.1274 x $10^{-8}$ & -9.1274 x $10^{-10}$ \\ 
 200 & -1.3131 x $10^{-7}$ & -1.3131 x $10^{-9}$ \\ 
 350 & -1.7496 x $10^{-7}$ & -1.7496 x $10^{-9}$ \\ 
 500 & -2.0972 x $10^{-7}$ & -2.0972 x $10^{-9}$ \\ 
 \hline
\end{tabular}
\end{center}
\caption{Shift in radial frequency $\delta\Omega^r$ for different values of semilatus rectum $p$ and coupling constant $\mu$}
\label{Table:1}
\end{table}
In Table \ref{Table:1}, $\delta\Omega^r$ values are shown for two different values of coupling constant $\mu$. It is evident that if we decrease the value of $\mu$ by the order of $10$ then $\delta\Omega^r$ decreases by order $100$. Empirically we can say that $\delta\Omega^r$ is proportional to $\mu^2$. 
\begin{figure}[!ht]
  \centering
 \includegraphics[scale=0.8]{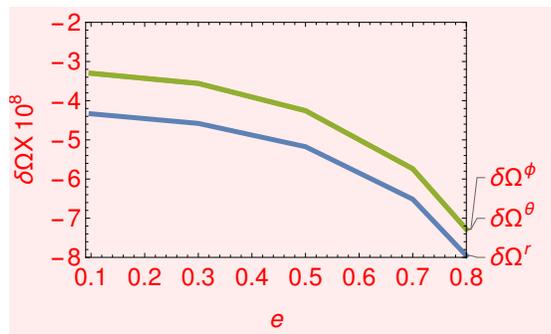}
\caption{$\Omega^i$ Vs. $e$ for $\mu=0.0001$, $p=30$ and $\theta_{min}=\pi/3$.}
\label{fig:orbitalfreq2} 
\end{figure}  

From Fig.[\ref{fig:orbitalfreq2}], it can be seen that the difference between $\delta\Omega^r$ and $\delta\Omega^{\theta}$ and $\delta\Omega^{\phi}$ decreases as the orbits become more eccentric. This result is in agreement with a result of \cite{Schmidt:2002qk} which says that the orbits having eccentricity near to unity are in Newtonian region and the orbital frequencies tend to be degenerate. 

According to canonical perturbation theory, the perturbed system has the same solution as the unperturbed one with frequency replaced by new frequency with shift\cite{classicalmec}. Thus, the deformed Kerr geometry has the same geodesic structure as the Kerr geometry with shifted frequencies as calculated by \eqref{obsfreqshift}.

 


\section{Acknowledgement} This work was partially supported by DST grant number SERB/PHY/2017041.




\end{document}